\documentclass[12pt]{article}

\usepackage{graphicx}
\usepackage{amssymb}
\usepackage{amsmath}
\usepackage{amsfonts}
\usepackage{amsthm}
\usepackage{verbatim}
\usepackage{tikz}
\usetikzlibrary{matrix,arrows}

\usepackage{comment}

\newtheorem{theorem}{Theorem}
\newtheorem{lemma}{Lemma}
\newtheorem{remark}{Remark}
\newtheorem{definition}{Definition}

\newtheorem{corollary}{Corollary}
\newtheorem{proposition}{Proposition}

\begin{document}


\title{Bases of T-meshes and the refinement of hierarchical B-splines}

\author{Dmitry Berdinsky \thanks{Department of Computer Science, The University of Auckland, Auckland 1142, New Zealand, e-mail: berdinsky@gmail.com}
\and
 Tae-wan Kim \thanks{Department of Naval Architecture and Ocean
 Engineering, Seoul National University, Seoul 151-744, Republic of
 Korea, e-mail: taewan@snu.ac.kr}
\and Durkbin Cho \thanks{Department of Mathematics, Dongguk University,
 Pil-dong 3-ga, Jung-gu, Seoul 100-715, Korea, e-mail: durkbin@dongguk.edu}
 \and Cesare Bracco \thanks{Department of Mathematics "Giuseppe Peano", University of
 Turin, v. Carlo Alberto 10, 10123, Turin, Italy, e-mail: cesare.bracco@unito.it}
 \and Sutipong Kiatpanichgij \thanks{School of Engineering and Technology, Asian Institute of Technology, Pathumthani 12120, Thailand, e-mail: s.kiatpanichgij@gmail.com}}

%
%

\date{}

\maketitle {\small
\begin{quote}
\noindent{ Abstract.   In this paper we consider spaces of bivariate splines of bi--degree $(m, n)$ with maximal order of smoothness over  domains associated to  a  two--dimensi\-onal grid. We define admissible classes of domains for which suitable  combinatorial technique allows us to obtain the     
  dimension  of such spline spaces and the number of tensor--product  B--splines acting effectively on these domains.   
     Following the strategy introduced recently by Giannelli and J\"{u}ttler, these results enable us to prove that under certain  
   assumptions about the configuration of a hierarchical T--mesh the hierarchical B-splines form a basis  of 
   bivariate splines of bi--degree $(m, n)$ with maximal order of smoothness over this hierarchical T--mesh. 
   In addition, we derive a sufficient condition about the 
   configuration of a hierarchical T-mesh that ensures a weighted partition of unity property for hierarchical B-splines with only positive   
   weights. } 

 \end{quote}

\section{Introduction}

  The spline representations using T-mesh as an underlying structure have
  absorbed substantial interest among designers for the last decade.
  The basic motivation to apply such representations in design
  and analysis is to break tensor-product structure of geometric
  representation used in NURBS. 
  A new interest in this issue has emerged
  recently in connection with isogeometric analysis, see Cottrell et al.~\cite{Hughes}.
  In this paper we deal with the concept of splines over T-meshes
  stated originally by Deng et al.~\cite{Deng05}.
  The issue of describing splines over a general T-mesh seems
  hardly solvable. In order to be able to generate
  spline basis functions and refine a spline space,
  we need to restrict ourselves on reasonable classes of T-meshes.

  For arbitrary TR-meshes, which include T-meshes, the dimension formula and  basis
  functions  have been derived for the $C^0$ case by Schumaker and Wang~\cite{Schumaker11}.
  For the case of bivariate splines of bi-degree $(m,n)$ with the reduced
  order of smoothness  $(r,r')$, i.e. $m \geqslant 2r+1 > 0$ and $n \geqslant 2r'+1 > 0$,
  spline basis functions have been obtained in terms of Bernstein-B\'ezier
  coefficients for T-meshes without cycles
  \cite{Schumaker12}. We note that this class of T-meshes
  includes the natural ones obtained
  as a result of refining a given rectangle by successive splitting
  rectangles into two subrectangles.
  For hierarchical T-meshes, the construction of PHT-splines~\cite{Deng06},
  which are splines of bi-degree $(3,3)$ and the order of smoothness
  $(1,1)$, showed an efficiency in surface modeling and
  isogeometric analysis.

  The construction of splines of bi-degree $(m,n)$ with the order of
  smoothness $(r,r')$ becomes more sophisticated for understanding when $m < 2r+1$ and $n <
  2r'+1$. It is worthwhile to analyze
  the class of hierarchical T-meshes for which the hierarchical B-splines,
  showing already their efficiency in applications, provide
  a basis of a spline space. Hierarchical B-splines for surface modeling were originally introduced by
  Forsey and Bartels~\cite{BartelsForsey88}. Kraft~\cite{Kraft}
  suggested a selection mechanism for hierarchical B-splines that
  ensures their linear independence as well as local refinement
  control. Vuong et
  al.~\cite{Vuong11} 
  considered applications of hierarchical B-splines in isogeometric analysis.

  Giannelli and J\"uttler \cite{Carlotta11} have recently proved
  that for a hierarchical T-mesh, determined by a nested sequence of
  domains $\Omega^0 \supset \dots  \supset \Omega^{N-1} \supset \Omega^{N} =
  \varnothing$ associated with a nested sequence of grids $G^0 \subset \dots \subset
  G^{N-1}$,  the hierarchical B-splines span the space of splines of bi-degree $(m,m)$ with maximal order of smoothness
  if each domain $R^{\ell} = \Omega^0 \setminus \Omega^\ell, \ell=1\dots N$, considered with respect to the grid $G^{\ell-1}$, lays in a certain  
  class.
  Later, that result has been generalised for splines of tri-degree  $(m,m,m)$  \cite{Berd13}. 
  
   In this paper we extend the results from \cite{Carlotta11} and \cite{Berd13} to the case of splines of bi--degree $(m,n)$
   with maximal order of smoothness.
   This extension requires new definition of admissible classes of domains associated with a two--dimensional grid. 
   We define these classes inductively and in a purely combinatorial way. 
   For a given bi--degree $(m,n)$, a two--dimensional grid and a domain from an admissible class, we obtain 
   the dimension of the spline space over this domain and the number of tensor--product B--splines acting effectively on it; 
   furthermore, it appears that these two numbers are equal.  
   Then, following the approach used in \cite{Carlotta11}, we prove that for certain assumptions about the 
   configuration of a hierarchical T--mesh the hierarchical B--splines form a basis of bivariate 
   splines of bi--degree $(m, n)$ with maximal order of smoothness over this hierarchical T--mesh.    
   Also, we find an additional condition about the configuration of a hierarchical T--mesh that ensures a weighted partition of unity property for 
   hierarchical B-splines with only positive weights.

  The rest of this paper is organized as follows. In Section
  \ref{chapterd1} we consider the basic one-dimensional case and
  prove  propositions necessary for Section \ref{chapterd2} where the
  two-dimensional case is considered.
  For given integers $k_1,k_2 \geqslant 0$ in Subsection \ref{subsecdild2}
  we introduce the class $\mathcal{A}_{k_1,k_2}^2 $ of two-dimensional domains formed by the cells of an infinite two-dimensional grid.
  In Subsection \ref{subsecd2} we derive the dimension for the
  space of tensor-product splines of bi-degree $(m,n)$ with
  maximal order of smoothness defined on a
  domain from the class $\mathcal{A}_{m-1,n-1}^2$.
  In Subsection \ref{subsecbasisd2} we show that a basis of this
  space can be obtained as the set of tensor-product B-splines
  acting effectively on the domain.
  In Section \ref{hiersection}, with the tools obtained in Section \ref{chapterd2},
  we provide a condition on the configuration of a hierarchical
  T-mesh to guarantee that hierarchical B-splines span the space of
  splines of bi-degree $(m,n)$ over this T-mesh (see Theorem \ref{finalcarlotta}).
  In addition, in Corollary \ref{sumtoone} we present a condition on
  a hierarchical T-mesh ensuring the existence of a weighted partition of unity for hierarchical
  B-splines, with only positive weights.
  We conclude this paper with several remarks in Section \ref{remarksection}.

 \section{Univariate case}
 \label{chapterd1}

 Let $T'$ be an infinite one-dimensional grid.
 For the sake of simplicity, we suppose that the distances between adjacent grid
 nodes of $T'$ are equal to $1$. 
 A cell of $T'$ is a closed segment
 of a length $1$ between adjacent grid nodes. Let $T'_{1}$ be the grid
 that is obtained by shifting $T'$ by $\frac{1}{2}$.

Let $\Omega$ be a closed bounded domain formed by a finite number of
cells of $T'$. 
Then, $\Omega$ consists of a number of segments of finite length. A
vertex of a domain $\Omega$ is a grid node of $T'$ that belongs to
$\Omega$. We say that a vertex of $\Omega$ is an inner vertex if it
lies in the interior of $\Omega$, which is hereinafter denoted by
$\mathrm{int}\,\Omega$. For a given domain $\Omega$ we define the
dilatation domains $\Omega^e_k$ in a recursive manner:
 \begin{definition}
\label{d1defdilatation}
  If $k=0$, $\Omega_0^e : = \Omega$.
  If $0 < k $ is odd, $\Omega_k^e$ is the union of the cells of $T'_1$
  with vertices of $\Omega_{k-1}^e$ as their centroids.
  If $0 < k $ is even, $\Omega_k^e$ is the union of the cells of $T'$
  with vertices of $\Omega_{k-1}^e$ as their centroids.
\end{definition}

 \begin{figure}[htp] \centering
  \includegraphics[width=1.0 \textwidth]{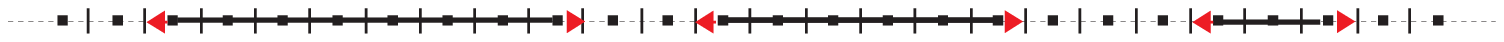}
  \caption{The grid nodes of $T'$ are denoted by black squares.
  The connected components of a domain $\Omega$ are denoted by thick solid lines.
  The grid nodes of $T'_1$ are denoted by the small vertical
  segments. The domain $\Omega_1 ^e$ is an union of $\Omega$
  and the adjacent segments of the length $\frac{1}{2}$ denoted by the triangles.
  We note that $\Omega \in \mathcal{A}_2 ^1 $, but $\Omega \notin \mathcal{A}_3 ^1$.}
  \label{odnomernayalinija}
\end{figure}

 An example of a domain $\Omega$ and its dilatation $\Omega_1 ^e$ are
 shown in Fig.~\ref{odnomernayalinija}. We observe that $\Omega_{k}^e \subset
 \Omega_{k+1}^e$ for any integer $k \geqslant 0$.
 For a given integer $k \geqslant
 0$, the class $\mathcal{A}_k ^1$ of one-dimensional domains is
 defined as follows:

\begin{definition}
\label{d1defoffset}
  We say that a domain $\Omega$ admits an offset at a distance of $\frac{k}{2}$
  if the number of cells between any two neighboring segments exceeds $k$.
  We denote by $\mathcal{A}_{k}^1$ the class of one-dimensional bounded domains
  that admit an offset at a distance of $\frac{k}{2}$.
\end{definition}
  We observe that $\mathcal{A}_{0}^1 \supset \mathcal{A}_{1}^1  \supset \mathcal{A}_{2}^1 \supset \dots$, and
  the class $\mathcal{A}_{0}^1$ includes all possible one-dimensional bounded domains formed by
 cells of $T'$.
 It can be seen that if  $\Omega \in \mathcal{A}_n ^1$, then $\Omega_k ^e$ has the same number of connected
 components for any $0 \leqslant k \leqslant n $.
 Propositions \ref{d1intersectionprop1} and \ref{d1intersectionprop2} below will be used in the proof of
 Proposition \ref{sectiondomd2} in Subsection \ref{subsecdild2}.
 \begin{proposition}
\label{d1intersectionprop1}
   Suppose that for domains $ \Omega_1 , \Omega_2 \in \mathcal{A}_0 ^1$,
   both intersection domains  $\Omega_1 \cap \Omega_2$ and $(\Omega_1)_1
   ^e  \cap (\Omega_2)_1 ^e$ belong to $ \mathcal{A}_0 ^1$.
   Then, $(\Omega_1)_1 ^e \cap (\Omega_2)_1 ^e  = (\Omega_1 \cap \Omega_2)_1 ^e$.
\end{proposition}
\noindent {\it Proof.} The inclusion $(\Omega_1 \cap \Omega_2)_1 ^e
\subset (\Omega_1)_1 ^e \cap (\Omega_2)_1 ^e $ is obvious. Let $C
\subset (\Omega_1)_1 ^e \cap (\Omega_2)_1 ^e $ be the cell $C$ of
$T'_1$  centered at a node $v$ of $T'$. Then, $v \in \Omega_1 \cap
\Omega_2$ and $C \subset (\Omega_1 \cap \Omega_2)^1 _e$.
 Since $(\Omega_1)_1 ^e \cap (\Omega_2)_1 ^e \in \mathcal{A}_0
 ^1$, it is composed of cells of $T'_1$ and, thus, the inclusion
 $(\Omega_1)_1 ^e \cap (\Omega_2)_1 ^e \subset (\Omega_1 \cap \Omega_2)_1 ^e
 $ is proved. $\Box$
\begin{proposition}
\label{d1intersectionprop2}
   Suppose that for domains $\Omega_1, \Omega_2 \in  \mathcal{A}_1
   ^1$, the intersection domain $\Omega_1 \cap \Omega_2$  belongs to $\mathcal{A}_0
   ^1$. Then, $\Omega_1 \cap \Omega_2 \in \mathcal{A}_1 ^1$.
\end{proposition}
\noindent {\it Proof.} Suppose that
 $\Omega_1 \cap \Omega_2 \notin \mathcal{A}_1 ^1$, then there exist
 neighboring segments of $\Omega_1 \cap \Omega_2$ with one cell between them.
 Then, at least either $\Omega_1$ or $\Omega_2$
 does not belong to $\mathcal{A}_1 ^1$. Thus, we have a
 contradiction. $\Box$

 Propositions \ref{d1cupprop1} and \ref{d1nodeprop2} below will be used in the proof of
 Proposition \ref{newlined2prop} in Subsection \ref{subsecdild2}.
 \begin{proposition}
 \label{d1cupprop1}
   Let $\Omega_1, \Omega_2 \in \mathcal{A}_0 ^1$. Then,
   \begin{equation}
   \label{d1cupprop1eq}
   (\Omega_1 \cup \Omega_2)_k ^e = (\Omega_1) _k ^e \cup (\Omega_2)_k ^e,
   \end{equation}
   for any integer $k \geqslant 0$.
 \end{proposition}
 \noindent {\it Proof:}
  For $k=0$ the identity \eqref{d1cupprop1eq} is trivial.
  Let us prove \eqref{d1cupprop1eq} for $k=1$. Let $C \subset (\Omega_1 \cup \Omega_2)_1
  ^e$ be the cell of $T_1 '$ centered at a vertex of $\Omega_1 \cup
  \Omega_2$. Then, $v$ is a vertex either of $\Omega_1$ or $\Omega_2$,
  i.e. either $C \subset (\Omega_1)_1 ^e $ or $C \subset (\Omega_2)_1
  ^e$. The inclusion $(\Omega_1 \cup  \Omega_2)_1 ^e \subset (\Omega_1)_1 ^e  \cup (\Omega_2)_1
  ^e$ is proved. The inclusion $(\Omega_1)_1 ^e  \cup (\Omega_2)_1
  ^e \subset (\Omega_1 \cup  \Omega_2)_1 ^e$ is verified in a
  similar way. Suppose that \eqref{d1cupprop1eq} is proved for some
  $k \geqslant 1$, then $(\Omega_1 \cup \Omega_2)_{k+1} ^e = ((\Omega_1 \cup \Omega_2)_k ^e )_1 ^e = ((\Omega_1)_k ^e \cup (\Omega_2)_k ^e)_1 ^e =
  ((\Omega_1)_k ^e)_1 ^e \cup ((\Omega_2)_k ^e)_1 ^e  = (\Omega_1)_{k+1} ^e \cup (\Omega_2)_{k+1}
  ^e$. $\Box$
   
   We note that Definition \ref{d1defoffset} does not depend on the distances between grid nodes of $T'$  
   and the classes $\mathcal{A}_k ^1,k \geqslant 0$ can be defined for an arbitrary one--dimensional grid $T'$.
   Hereinafter in this section, we will no longer suppose that the distances between adjacent grid nodes are equal to $1$.   
   \begin{proposition}
 \label{d1nodeprop2}
  For a given integer $k \geqslant 0$, let  $\Omega \in \mathcal{A}_k ^1$. Let us add a new node $v$ to the grid $T'$. Then,
  a domain $\Omega$, considered with respect to the grid $T' \cup \{ v \}$, belongs to the class $\mathcal{A}_k ^1$ as well.
 \end{proposition}
 \noindent {\it Proof:}
 If one add a node to the grid  $T'$, then a number of cells between
 two neighboring segments either does not change or increases in one.
 Thus, the proposition is proved.$\Box$

Let $R_m$ be the vector space of univariate polynomials of degree
$m$. Let $\mathcal{T}$ be a mesh, which is a portion of  $T'$
included in a domain $\Omega$. We denote by $\mathcal{S}_m
(\mathcal{T})$ the vector space of $C^{m-1}$ smooth functions
defined on $\Omega$ that are polynomials in $R_m$ on each cell of a
domain $\Omega$. We denote by $f_1$ and $f_0 ^0$ the number of cells
forming a domain $\Omega$ and the number of inner vertices of
$\Omega$, respectively.
\begin{proposition}
\label{dimd1}
  For a given domain $\Omega$, the dimension of the corresponding
  spline space is:
  \begin{equation*}
    \mathrm{dim}\, \mathcal{S}_m (\mathcal{T}) =
    (m+1) f_1 -  m f_0 ^0.
  \end{equation*}
\end{proposition}
\noindent {\it Proof.} On each cell of $\mathcal{T}$ the spline
function of $\mathcal{S}_m (\mathcal{T})$ is a polynomial of degree
$m$, and for each inner vertex there are $m$ linearly independent constraints on
the coefficients of these polynomials. Thus, the proposition is proved.$\Box$

For a given integer $m \geqslant 1$, let $\mathcal{B}$ be the set of
segments formed by $m+1$ consecutive cells of $T'$, so $\mathcal{B}$
is the set of all possible minimal supports for B-splines of degree
$m$ defined over $T'$ and with knot multiplicities equal to $1$. We
denote by $\widehat{\mathcal{B}}$ the collection of B-splines whose
supports become the elements of $\mathcal{B}$. Let $\mathcal{N}$ be
the number of elements of $\mathcal{B}$ that have at least one cell
in common with a domain $\Omega$. 

\begin{proposition}
\label{Nd1}
  For a given integer $m \geqslant 1$, suppose that $\Omega \in \mathcal{A}_{m-1} ^1$.
  Then, the following identity holds:
  \begin{equation}
  \label{d1N}
    \mathcal{N} = (m+1) f_1  - m f_0 ^0.
  \end{equation}
\end{proposition}
\noindent {\it Proof.} Suppose that $\Omega$ has one connected
component. Then, the simple observations that: $\mathcal{N} = m +
f_1 $ and $f_1 - f_0 ^0= 1$ prove \eqref{d1N}. Since $\Omega \in
\mathcal{A}_{m-1} ^1$, each element of $\mathcal{B}$ may have cells
in common with no more than one connected component of $\Omega$.
Thus, the identity \eqref{d1N} holds for any $\Omega \in
\mathcal{A}_{m-1} ^1$. $\Box$

\begin{corollary}
\label{maintheoremd1}
  For a given $m \geqslant 1$, suppose that $\Omega \in \mathcal{A}_{m-1}^1$.
  Then, the basis of a space $\mathcal{S}_m (\mathcal{T})$ can be obtained
  as follows:
  \begin{equation}
    \{ b|_{\Omega} : b (x) \in \widehat{\mathcal{B}} \,\, \wedge \,\,
    \mathrm{supp} \, b (x) \cap \mathrm{int} \, \Omega \neq \varnothing\}.
  \end{equation}
\end{corollary}
\noindent{\it Proof.}  Corollary \ref{maintheoremd1} is a direct
consequence  of Propositions~\ref{dimd1} and~\ref{Nd1}, and the fact
that the one-dimensional B-splines from $\widehat{\mathcal{B}}$ are
linear independent. $\Box$

\begin{corollary}
\label{extensiond1}
  For a given integer $m \geqslant 1$, suppose that $\Omega \in \mathcal{A}_{m-1} ^1$.
  Let $f \in \mathcal{S}_m (\mathcal{T})$ be a spline function defined over
  the corresponding mesh $\mathcal{T}$.
  Then, there exists a spline function $\widetilde{f}$ of degree $m$
  defined globally over $T'$ such that $\widetilde{f}|_\Omega =  f$.
\end{corollary}
\noindent {\it Proof.} The B-splines from  $\widehat{\mathcal{B}}$
are defined globally over $T'$. Thus, by
Corollary~\ref{maintheoremd1}, the corollary is proved. $\Box$

\section{Bivariate case}
\label{chapterd2}

  Let $T'$ be a two-dimensional infinite grid.
  For the sake of simplicity, we suppose that the distances
  between adjacent grid nodes of $T'$ are equal to $1$. A cell of $T'$
  is a closed square with sides of length $1$ aligned with the grid
  lines of $T'$.

  Let $\Omega$ be a closed bounded domain formed by cells of $T'$
  (for example see the diagonally hatched area shown in Fig.~\ref{figd2book2h}).
  A vertex of a domain $\Omega$  is a grid node of $T'$ that belongs to $\Omega$.
  We say that a vertex is a boundary vertex if it belongs to $\partial\Omega$, and
  we say that a vertex is an inner vertex if it lies in the interior
  of $\Omega$. An edge of a domain $\Omega$ is a closed segment
  between two adjacent grid nodes of $T'$, which is a subset of
  $\Omega$. We say that an edge is a boundary edge if it is a subset
  of $\partial\Omega$, and we say that an edge is an inner edge if it
  is not a boundary edge.
  Throughout this section we will suppose that $\Omega$ is a two-dimensional
  topological manifold with a boundary. A violation of this
  restriction can occur only in a neighborhood of a boundary
  vertex. The admissible and inadmissible configurations for a neighborhood of
  a boundary vertex are shown in Fig.~\ref{figd2book1e}.
  Additionally, we remark that a domain $\Omega$ may have several connected components.
  \begin{figure}[htp] \centering
  \includegraphics[width=0.5 \textwidth]{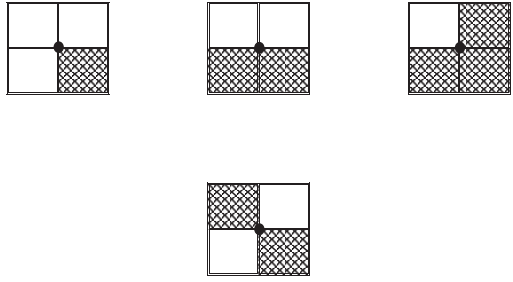}
  \caption{The boundary vertices, which are shown with black dots,  are at the centroids of $2 \times 2$ squares
  that are formed by four cells of $T'$.
  The diagonally hatched cells belong to $\Omega$.
  Three admissible configurations are shown at the top.
  The inadmissible configuration is shown at the bottom.}
  \label{figd2book1e}
  \end{figure}

\subsection{Vertical and horizontal dilatations of a two-dimensional domain}
 \label{subsecdild2}
 Let $T'_{1,0}$, $T'_{0,1}$ and $T'_{1,1}$ be the grids that are obtained by shifting $T'$ by the
 vectors $\left( \frac{1}{2}, 0 \right)$, $\left( 0, \frac{1}{2}\right)$ and $\left( \frac{1}{2}, \frac{1}{2}\right)$,
 respectively. For the grid $T'$ we will also use the notation $T'_{0,0}:=T'$.
 For a given domain $\Omega$ and integers $k_1, k_2 \geqslant 0$, we define the dilatation domains
 $\Omega_{k_1,k_2}^e$ in a recursive manner:

\begin{figure}[htp] \centering
  \includegraphics[width=0.51 \textwidth]{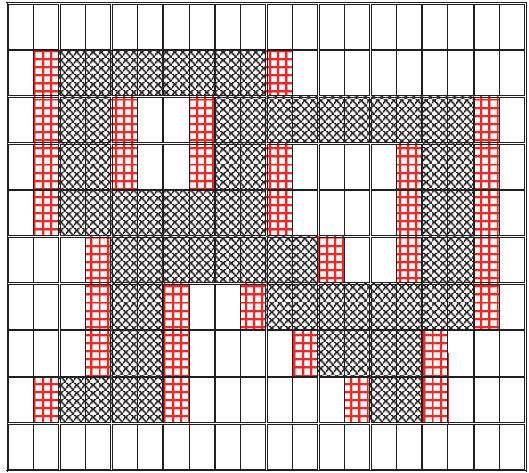}
  \caption{The grid $T'$ is aligned with thick solid lines.
  The cells of a domain $\Omega$ are diagonally hatched.
  The grid $T'_{1,0}$, which is shifted by the vector $\left( \frac{1}{2}, 0 \right)$ from $T'$,
  is aligned with thin solid vertical lines and thick solid horizontal lines.
  The domain $\Omega_{1,0}^e$ is the whole shaded area and it is formed
  by the cells of $T'_{1,0}$.}
  \label{figd2book2h}
  \end{figure}

  \begin{figure}[htp] \centering
 \includegraphics[width=0.51 \textwidth]{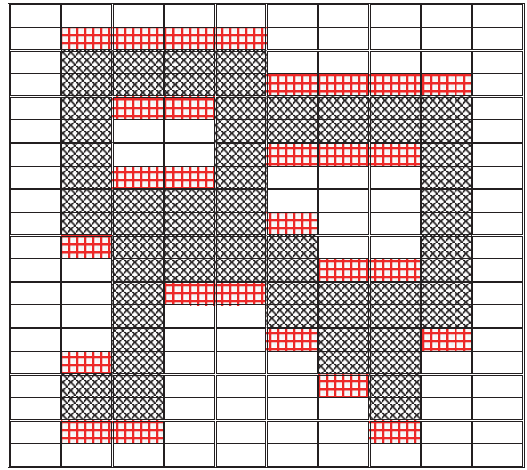}
 \caption{The grid $T'$ is aligned with thick solid lines.
 The cells of a domain $\Omega$ are diagonally hatched.
 The grid $T'_{0,1}$, which is shifted by the vector $\left( 0, \frac{1}{2} \right)$ from $T'$,
 is aligned with thick solid vertical lines and thin solid horizontal lines.
 The domain $\Omega_{0,1}^e$ is the whole shaded area and it is formed
 by the cells of $T'_{0,1}$.}
 \label{figd2book2v}
 \end{figure}

 \begin{definition}
 \label{d2defdilatation}
 If $k_1=k_2=0$, $\Omega_{0,0}^e := \Omega$.
 If $k_1 = 1$ and $k_2 = 0$, $\Omega_{1,0} ^e$ is the union of
 cells of $T'_{1,0}$ that are split into two equal halves by vertical edges of $\Omega$.
 An example of horizontal dilatation is shown in
 Fig.~\ref{figd2book2h}.
 If $k_1 = 0$ and $k_2 = 1$, $\Omega_{0,1} ^e$ is the union of
 cells of $T'_{0,1}$ that are split into two equal halves by horizontal edges of $\Omega$.
 An example of vertical dilatation is shown in
 Fig.~\ref{figd2book2v}.
 If $k_1 + k_2 > 1$, then the following recursion can be defined in both ways:
 \begin{equation}
 \label{dilatationd2}
 \Omega_{k_1 , k_2} ^e : = \left( \Omega_{k_1 - 1, k_2} ^e \right)_{1,0} ^e = \left( \Omega_{k_1 , k_2 -1} ^e \right)_{0,1} ^e.
 \end{equation}

 We remark that $\Omega_{k_1,k_2} ^e$ is formed by cells of $T'_{k_1 \, \mathrm{mod} \, 2, \, k_2 \, \mathrm{mod} \, 2}$.
 An example of the domain $\Omega_{1,1} ^e$ is shown in \cite{Berd13},
 Fig.~4 (where the infinite grid $T'_{1,1}$ is denoted as $T''$).
 \end{definition}

Let the array of cells of $T'$ be indexed by $\mathbb{Z} \times \mathbb{Z}$,
 i.e. each cell has index $(i,j)$ (where $i$ and $j$ are the indices of
 row and column containing this cell, respectively). Without loss of
 generality suppose that the lowest row and the leftmost column of this array that
 contain cells from $\Omega$ have index $1$.
 One may represent a domain $\Omega$ as a union 
 $\Omega = \bigcup_{i=1} ^{m_1} \Omega^h _i$, where each
 $\Omega^h _i$ is formed by cells of $\Omega$ in the $i$th row of cells;
 $m_1$ is the maximal index of row that contains cells from $\Omega$.
 Let $H_i, 1 \leqslant i \leqslant
 m_1$ be the projection of $\Omega^h _i$ onto horizontal grid line,
 i.e. $H_i \in \mathcal{A}_0 ^1$.
 Similarly, one may represent a domain $\Omega$ as a union
 $\Omega = \bigcup_{j=1} ^{m_2} \Omega^v _j$, where each
 $\Omega^v _j$ is formed by cells of $\Omega$ in the $j$th column of
 cells; $m_2$ is the maximal index of column that
 contains cells from $\Omega$. Let $V_j, 1 \leqslant j \leqslant m_2$ be
 the projection of $\Omega^v _j$ onto vertical grid line,
 i.e. $V_j \in \mathcal{A}_0 ^1$.
 We denote by $H$ and $V$ the maps that transform a domain $\Omega$
 to the ordered sequences of one-dimensional domains $\langle H_1, H_2, \dots, H_{m_1} \rangle$ and
 $\langle V_1, V_2, \dots, V_{m_2} \rangle$,
    respectively.  The following proposition is a trivial observation from  Definition \ref{d2defdilatation}.

  \begin{proposition}
  \label{d2HVext}
    Let   $\Omega = \bigcup_{i=1} ^{m_1}
    \Omega^h _i = \bigcup_{j=1} ^{m_2} \Omega^v  _j$ be a domain with the projections on horizontal and vertical
    grid-lines: $H_i, 1 \leqslant i \leqslant m_1$ and $V_j, 1 \leqslant j \leqslant m_2$, respectively (see Definition \ref{d2defoffset}).
    respectively.
    Then,
    \begin{equation}
     \label{d2HVextH}
      H(\Omega_{0,1} ^e) = \langle  H_1, H_1 \cup H_2, \dots, H_{m_1-1} \cup H_{m_1}, H_{m_1} \rangle = \sigma H (\Omega),
    \end{equation}
    \begin{equation}
    \label{d2HVextV}
          V(\Omega_{1,0} ^e) = \langle V_1, V_1 \cup V_2, \dots, V_{m_2 -1} \cup V_{m_2}, V_{m_2} \rangle = \sigma V(\Omega),
    \end{equation}
    where $\sigma$ is a map $\sigma:  \langle D_1, \dots, D_n \rangle \mapsto \langle D_1, D_1 \cup D_2, \dots, D_{n-1} \cup D_n , D_n
    \rangle$.
  \end{proposition}
 
 For  given integers $k_1, k_2 \geqslant 0$, we define the class $\mathcal{A}_{k_1,k_2}^2$ of two-dimensional domains as follows:

 \begin{definition}
 \label{d2defoffset}

 We denote by $\mathcal{A}_{0,0} ^2$ the class of all bounded two--dimensional
 domains formed by the cells of $T'$ that are topological manifolds with boundary.

 We say that a domain $\Omega$ admits horizontal offset at a distance of $\frac{1}{2}$
 if $H (\Omega) \cup \sigma H(\Omega) \subset \mathcal{A}_1 ^1$, i.e.
  each  intersection of $\Omega$ with  a horizontal line belongs to the class $\mathcal{A}_1 ^1$
   with respect to the one-dimensional grid, which is a horizontal projection of $T'$.
 We denote by $\mathcal{A}_{1,0} ^2$ the class of two--dimensional
 domains $\Omega \in \mathcal{A}_{0,0} ^2$ that admit horizontal
 offset at a distance of $\frac{1}{2}$.
 An example of a domain from the class $\mathcal{A}_{1,0} ^2$ is shown in
 Fig.~\ref{figd2book2h}.

 We say that a domain $\Omega$
 admits vertical offset at a distance of $\frac{1}{2}$
 if $V (\Omega) \cup \sigma V(\Omega) \subset \mathcal{A}_1 ^1$ , i.e.
 each  intersection of $\Omega$ with  a vertical line belongs to the class $\mathcal{A}_1 ^1$
 with respect to the one-dimensional grid, which is a vertical projection of $T'$.
 We denote by $\mathcal{A}_{0,1} ^2$ the class of two--dimensional
 domains $\Omega \in \mathcal{A}_{0,0} ^2$ that admit vertical
 offset at a distance of $\frac{1}{2}$. An example of a domain
 from the class $\mathcal{A}_{0,1} ^2$ is shown in
 Fig.~\ref{figd2book2v}.

 For a given integer $k \geqslant 1$  suppose that classes
 $\mathcal{A}_{k_1,k_2} ^2$ are defined for all
 nonnegative $k_1, k_2$ such that $k_1 + k_2 \leqslant k$.
 For nonnegative integers $k_1, k_2$ such that $k_1 + k_2 = k +1$,
 the class $\mathcal{A}_{k_1, k_2} ^2$ is defined as
 follows:
 \begin{enumerate}
 \item For $k_2 = 0$, $\Omega \in \mathcal{A}_{k+1, 0} ^2$ if
 $\Omega \in \mathcal{A}_{k , 0} ^2$ and
 $\Omega_{k, 0}^e \in \mathcal{A}_{1,0} ^2$.
 \item For $k_1 = 0$, $\Omega \in \mathcal{A}_{0, k+1} ^2$ if
 $\Omega \in \mathcal{A}_{0,k} ^2$ and
 $\Omega_{0,k}^e \in \mathcal{A}_{0,1} ^2$.
 \item For $k_1, k_2 > 0$, then the class $\mathcal{A}_{k_1, k_2} ^2$
 might be defined in two equivalent ways:
 \begin{enumerate}
 \item  $\Omega \in \mathcal{A}_{k_1, k_2} ^2$ if $\Omega \in \mathcal{A}_{k_1 -1, k_2} ^2
 $ and $\Omega_{k_1 -1 , k_2} ^e \in \mathcal{A}_{1,0} ^2$.
 \item  $\Omega \in \mathcal{A}_{k_1, k_2} ^2$ if $\Omega \in \mathcal{A}_{k_1 , k_2 -
 1} ^2 $ and $\Omega_{k_1, k_2 - 1} ^e \in \mathcal{A}_{0,1} ^2$.
 \end{enumerate}
 \end{enumerate}
  The equivalence of two ways to construct $\mathcal{A}_{k_1, k_2} ^2$
  given in the items 3(a) and 3(b) is proven in
  Corollary \ref{d2offsetdefeqiuv}.
 \end{definition}

  \begin{figure}[htp] \centering
 \includegraphics[width=0.2 \textwidth]{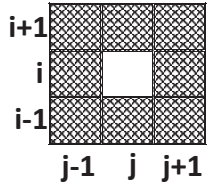}
 \caption{ The grid $T'$ is aligned with thick solid lines.
 The diagonally hatched cells belong to a domain $\widehat{\Omega}$.
 The indices $i-1,i,i+1$ and $j-1, j, j+1$ denote the corresponding rows and columns of $T'$, respectively.}
 \label{figd2ass1}
 \end{figure}

  \begin{corollary}
  \label{d2offsetdefeqiuv}
  For given integers $k_1, k_2 > 0$ , the following identity between
  two sets of domains holds:
  \begin{equation}
  \label{idensetd2}
     \{ \Omega \in \mathcal{A}_{k_1-1 , k_2} ^2| \Omega_{k_1 -1 , k_2 } ^e
     \in \mathcal{A}_{1,0} ^2 \}  =
     \{ \Omega \in \mathcal{A}_{k_1 , k_2 -1} ^2| \Omega_{k_1 , k_2 -1 } ^e
     \in \mathcal{A}_{0,1} ^2 \}.
   \end{equation}
  \end{corollary}
  \noindent {\it Proof.}
  Let us denote the left and right parts of \eqref{idensetd2} as
  $L_{k_1,k_2}$ and $R_{k_1, k_2}$, respectively. Then,
  \begin{equation*}
     L_{k_1,k_2} = \{ \Omega \in \mathcal{A}_{k_1-1 , k_2 - 1} ^2 | \Omega_{k_1 -1 , k_2 -1 } ^e
     \in \mathcal{A}_{0,1} ^2 \, \wedge \, \Omega_{k_1 -1 , k_2 } ^e
     \in \mathcal{A}_{1,0} ^2  \},
    \end{equation*}
     \begin{equation*}
     R_{k_1, k_2} = \{ \Omega \in \mathcal{A}_{k_1 -1 , k_2 -1} ^2 | \Omega_{k_1 -1 , k_2 -1 } ^e
     \in \mathcal{A}_{1,0} ^2 \, \wedge \, \Omega_{k_1 , k_2 -1 } ^e
     \in \mathcal{A}_{0,1} ^2 \}.
     \end{equation*}
 Let us prove the inclusion: $L_{k_1, k_2} \subset R_{k_1 ,k_2}$.
 For a given $\Omega \in L_{k_1, k_2}$, $\widehat{\Omega} := \Omega_{k_1 -1 , k_2 -1 }^e \in \mathcal{A}_{0,1} ^2  $ and
 $\widehat{\Omega}_{0,1} ^e \in \mathcal{A}_{1,0} ^2$. From
 Proposition \ref{d2HVext}  
  we get that $V(\widehat{\Omega}) \cup \sigma V(\widehat{\Omega}) \subset \mathcal{A}_1 ^1 $
 and $\sigma H (\widehat{\Omega}) \cup \sigma ^2 H (\widehat{\Omega}) \subset \mathcal{A}_1 ^1$.

 Assume that $H (\widehat{\Omega}) \not \subset \mathcal{A}_1 ^1$,
 i.e. there exists $\widehat{H}_i \in H (\widehat{\Omega})$ for some $i$ such that $\widehat{H}_i \notin \mathcal{A}_1 ^1 $.
 Since $\sigma H (\widehat{\Omega}) \subset \mathcal{A}_1 ^1$, then
 $\widehat{H}_{i-1},\widehat{H}_{i+1} \in H(\widehat{\Omega})$ are non-empty, and $\widehat{H}_i \cup \widehat{H}_{i-1}  \in \mathcal{A}_1
 ^1$, $\widehat{H}_i \cup \widehat{H}_{i+1}  \in \mathcal{A}_1
 ^1$. An assumption $\widehat{H}_i \notin \mathcal{A}_1 ^1$ means
 that there exist neighboring segments of $\widehat{H}_i$ with a distance
 equal to one between them. Thus, we have the following local
 picture of the domain $\widehat{\Omega}$ in Fig.~\ref{figd2ass1}.
 From Fig.~\ref{figd2ass1} we see that for $\widehat{V}_j \in V (\widehat{\Omega})$,
 $\widehat{V}_j \notin \mathcal{A}_1 ^1$. Since $V (\widehat{\Omega}) \subset \mathcal{A}_1 ^1
 $, we have a contradiction.

  \begin{figure}[htp] \centering
  \includegraphics[width=0.2 \textwidth]{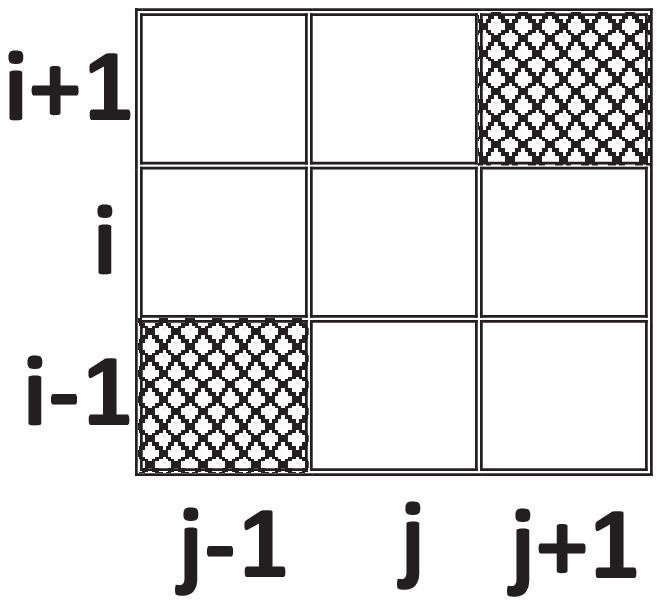}
  \caption{ The grid $T'$ is aligned with thick solid lines.
  The diagonally hatched cells belong to a domain $\widehat{\Omega}$.
  The indices $i-1,i,i+1$ and $j-1, j ,j+1$ denote the corresponding rows and columns of $T'$, respectively.}
  \label{figd2ass2}
  \end{figure}

  Suppose that $\sigma ^2 V (\widehat{\Omega}) \not \subset \mathcal{A}_1
  ^1$, i.e. there exist $\widehat{V}_{j-1}, \widehat{V}_{j},
  \widehat{V}_{j+1} \in V (\widehat{\Omega})$ for some $j$ such that
  $\widehat{V}_{j-1} \cup \widehat{V}_j \cup \widehat{V}_{j+1} \notin \mathcal{A}_1
  ^1$. Since $V (\widehat{\Omega}) \cup \sigma V (\widehat{\Omega}) \subset \mathcal{A}_1 ^1
  $, then $V_{j-1}, V_j, V_{j+1}, V_{j-1} \cup V_j , V_{j+1} \cup V_j \in \mathcal{A}_1
  ^1$. An assumption $\widehat{V}_{j-1} \cup \widehat{V}_j \cup \widehat{V}_{j+1} \notin \mathcal{A}_1
  ^1$ means that there are neighboring segments of $\widehat{V}_{j-1} \cup \widehat{V}_j \cup
  \widehat{V}_{j+1}$ with a distance equal to one between them.
  Thus, we have the following local picture of the domain
  $\widehat{\Omega}$ in Fig.~\ref{figd2ass2}.
  From Fig.~\ref{figd2ass2} we see that for $\widehat{H}_{i-1} \cup \widehat{H}_i \cup \widehat{H}_{i+1}
  \in \sigma^2  H (\widehat{\Omega})$, $\widehat{H}_{i-1} \cup \widehat{H}_i \cup \widehat{H}_{i+1} \notin \mathcal{A}_1
  ^1$. Since $\sigma ^2 H ( \widehat{\Omega}) \subset \mathcal{A}_1 ^1$, we have a
  contradiction.  By contradiction we have proved that $H (\widehat{\Omega})\cup \sigma^2 V (\widehat{\Omega}) \subset \mathcal{A}_1
  ^1$. Therefore, $H (\widehat{\Omega}) \cup \sigma H (\widehat{ \Omega}) \subset \mathcal{A}_1 ^1
  $ and $ \sigma V (\widehat{\Omega}) \cup \sigma^2 V (\widehat{\Omega}) \subset \mathcal{A}_1
  ^1$, which imply that $\widehat{\Omega} \in \mathcal{A}_{1,0} ^2$
  and $\widehat{\Omega}_{1,0} ^e \in \mathcal{A}_{0,1} ^2$, and,  thus, $\Omega \in R_{k_1,
  k_2}$. The inclusion $L_{k_1,k_2} \subset R_{k_1,k_2}$ is proven.
  The reverse inclusion could be proven in an analogous way.

  In order to prove correctness of Definition \ref{d2defoffset}, namely an equivalence of the items 3(a) and 3(b),  one
  need to apply the arguments above to the basic case $k_1 = k_2 =1$
  and then proceed by induction on $k_1 + k_2$.$\Box$

  Proposition \ref{sectiondomd2} below will be
  used in the proof of Corollary \ref{cordimd2} in  Subsection \ref{subsecd2}.

  \begin{proposition}
  \label{sectiondomd2}
     Let $\Omega  =  \bigcup_{i=1} ^{m_1} \Omega^h _i =   \bigcup_{j=1} ^{m_2} \Omega^v _i  \in \mathcal{A}_{k_1,
     k_2} ^2$. For some $i,j$ such that $1 \leqslant i \leqslant m_1 -1, 1 \leqslant j \leqslant m_2 -
     1$, let $H_i, H_{i+1} \in H (\Omega)$ and $V_j, V_{j+1} \in V(\Omega)$ be the corresponding
     projections of $\Omega^h _i, \Omega^h _{i+1}$ and $\Omega^v _j, \Omega^v _{j+1}$
     respectively. Then, for one-dimensional domains
     $H_i \cap H_{i+1} = \Omega^h _i \cap \Omega^h _{i+1}$ and
     $V_{j} \cap V_{j+1} = \Omega^v _j \cap \Omega^v _{j+1}$,
     $H_i \cap H_{i+1} \in \mathcal{A}_{k_1} ^1$ and $V_{j} \cap V_{j+1} \in \mathcal{A}_{k_2} ^1$.
  \end{proposition}
  \noindent {\it Proof.}
  It follows from Definition \ref{d2defoffset} that $\Omega \in \mathcal{A}_{k_1,
  0} ^2$. Let us prove that $H_i \cap H_{i+1} \in \mathcal{A}_{k_1}
  ^1$.
  Firstly, we note that
  $H_i \cap H_{i+1} = \Omega^h _{i} \cap \Omega^h _{i+1} \in \mathcal{A}_0
  ^1$, because $\Omega$ is a topological manifold with boundary.
  Secondly, one can see that Definition \ref{d1defoffset} might be given
  by induction, similarly to Definition \ref{d2defoffset},
  namely: $D \in \mathcal{A}_{k+1} ^1$ if $D \in \mathcal{A}_{k} ^1$ and $D_k ^e \in \mathcal{A}_1 ^1 $ for $k \geqslant 1$.
  Thus, from Propositions \ref{d1intersectionprop1}, \ref{d1intersectionprop2} and Definition \ref{d2defoffset} it follows
  that:
  $$
  (H_{i})_{l} ^e  \cap (H_{i+1})_{l} ^e  = (H_i \cap H_{i+1})_{l} ^e,
  $$
  for any $l \leqslant k_1$,  and $H_{i+1} \cap H_i \in \mathcal{A}_{k_1}
  ^1$. In an analogous way one can prove that $V_{j} \cap V_{j+1} \in \mathcal{A}_{k_2} ^1$.$\Box$

  We note that Definition \ref{d2defoffset} does not depend on the distances between grid nodes of $T'$  
   and the classes $\mathcal{A}_k ^2,k \geqslant 0$ can be defined for an arbitrary two--dimensional grid $T'$.
   Hereinafter in this section, we will no longer suppose that the distances between adjacent grid nodes are equal to $1$.   
 Proposition \ref{newlined2prop} below will be used in the proof of Proposition
 \ref{refhierprop1} in Section \ref{hiersection}.
 \begin{proposition}
 \label{newlined2prop}
  For given integers $k_1, k_2 \geqslant 0 $, let $\Omega \in
  \mathcal{A}_{k_1,k_2} ^2$. Let us add new vertical or horizontal line $l_1$ to the
  grid $T'$. Then, a domain $\Omega$, considered with respect to the
  grid $T' \cup \{l_1 \}$, belongs to the class $\mathcal{A}_{k_1,k_2} ^2$ as well.
 \end{proposition}
 \noindent {\it Proof.} Suppose that $l_1$ is vertical line. Since $\Omega \in \mathcal{A}_{k_1,k_2} ^2$, then $\Omega \in \mathcal{A}_{0,k_2}^2$.
  Then, it can be seen from Definition \ref{d2defoffset} that $\Omega$,
  considered with respect to the grid $T' \cup \{l_1 \}$, belongs to the class
  $\mathcal{A}_{0,k_2}^2$ as well.
  In addition, we observe that a domain $\Omega_{0,k_2}  ^e$ does
  not depend on whether the vertical dilatation of $\Omega$ is considered
  with respect to $T'$ or $T' \cup {l_1}$.
  Since $\Omega \in \mathcal{A}_{k_1,k_2} ^2$, then $\Omega_{0,k_2} ^e \in \mathcal{A}_{k_1,0}^2$.
  It follows straightforwardly from Propositions \ref{d1cupprop1} and \ref{d1nodeprop2}
     that $\Omega_{0,k_2} ^e$, considered with respect to the grid $T' \cup l_1$,
  belongs to the class $\mathcal{A}_{k_1,0} ^2$ as well. Similarly, one can prove Proposition \ref{newlined2prop} for a horizontal line $l_1$.
  $\Box$

 \begin{corollary}
 \label{newlined2}
   For given integers $k_1, k_2 \geqslant 0 $, let $\Omega \in
  \mathcal{A}_{k_1,k_2} ^2$. Let us add a finite number of new vertical and horizontal lines $l_1, \dots, l_k$ to the
  grid $T'$. Then, the domain $\Omega$, considered with respect to the
  grid $T' \cup \{ l_1,  \dots, l_k \}$, belongs to the class $\mathcal{A}_{k_1,k_2} ^2$ as well.
 \end{corollary}
 \noindent{\it Proof:} Corollary \ref{newlined2} is a straightforward consequence of Proposition \ref{newlined2prop}.$\Box$

 Definition \ref{d2inneroffset} below will be needed to state
 Propositions \ref{finalcarlottaprop} and \ref{finalteor2} in Section \ref{hiersection}.
 \begin{definition}
  \label{d2inneroffset}
   For a given $\Omega \in \mathcal{A}_{0,0} ^2$, let $\widehat{\Omega}$
   be a sufficiently large rectangle formed by the cells of $T'$ such
   that $\Omega \subset  \mathrm{int} \, \widehat{\Omega}$.
   For given integers $k_1,k_2 \geqslant 0$, we say
   that $\Omega \in \widetilde{\mathcal{A}}_{k_1,k_2} ^2$ if a domain $\widehat{\Omega} \setminus \mathrm{int}\, \Omega
   $ belongs to  $ \mathcal{A}_{k_1,k_2}^2$. We remark that $\widetilde{\mathcal{A}}_{0,0} ^2 =
   \mathcal{A}_{0,0}^2$.
 \end{definition}

 \begin{corollary}
 \label{newlined2inn}
   For given integers $k_1, k_2 \geqslant 0 $, let $\Omega \in
  \widetilde{\mathcal{A}}_{k_1,k_2} ^2$. Let us add a finite number of new vertical and horizontal lines $l_1, \dots, l_k$ to the
  grid $T'$. Then, the domain $\Omega$, considered with respect to the
  grid $T' \cup \{ l_1,  \dots, l_k \}$, belongs to the class $\widetilde{\mathcal{A}}_{k_1,k_2} ^2$ as well.
 \end{corollary}
 \noindent {\it Proof.} Corollary \ref{newlined2inn} is a straightforward consequence of Definition \ref{d2inneroffset} and
 Corollary \ref{newlined2}.$\Box$

\subsection{Dimension of a spline space over a two-dimensional domain}
\label{subsecd2}

  Let $R_{m,n}$ be the vector space of polynomials of bi-degree $(m,n)$
  with respect to two variables $x$ and $y$. Let $\mathcal{T}$ be a
  T-mesh
  , which is a portion of $T'$ included in a
  domain $\Omega$. We denote by $\mathcal{S}_{m,n} (\mathcal{T})$ the
  vector space of $C^{m-1, n-1}$ smooth functions defined on $\Omega$ that
  are polynomials in $R_{m,n}$ on each cell of a domain $\Omega$.
  We denote by $f_2$, $f_1 ^{h,0}$, $f_1 ^{v,0}$ and $f_0 ^0$  the numbers of cells, horizontal inner
  edges, vertical inner edges and inner vertices of a domain $\Omega$, respectively.
\begin{proposition}[Mourrain~\cite{Mourrain12}, Theorem~3.3 and Corollary~3.2]
\label{cordimd2noholes}
  Suppose that a domain $\Omega$ corresponding to the T-mesh $\mathcal{T}$
  is simply connected.
  Then,
  \begin{equation}
  \label{dimd2equ}
  \begin{split}
    \mathrm{dim}\, \mathcal{S}_{m,n} (\mathcal{T}) =
    (m+1) (n+1) f_2 -  ( (m+1) n  f_1 ^{h,0}  + \\
    (n+1) m f_1 ^{v,0} ) + m n f_0 ^0.
   \end{split}
  \end{equation}
\end{proposition}

In the following lemma we will obtain the dimension of a spline
space $\mathrm{dim}\, \mathcal{S}_{m,n} (\mathcal{T})$ if the
corresponding domain $\Omega$ is split into two domains $\Omega_1$
and $\Omega_2$ (see Fig.~\ref{figd2book2cut}). Let $U'$ be a  grid
line of $T'$. We say that $U'$ splits a domain $\Omega$ into two
nonempty domains if $\Omega = \Omega_1 \cup \Omega_2$, where
$\Omega_1$ and $\Omega_2$ are contained in different half-spaces
divided by $U'$. We denote by $U$ the corresponding one-dimensional
domain $U = \Omega_1 \cap \Omega_2$ formed by one-dimensional cells
of $U'$.

 \begin{figure}[htp] \centering
  \includegraphics[width=0.4 \textwidth]{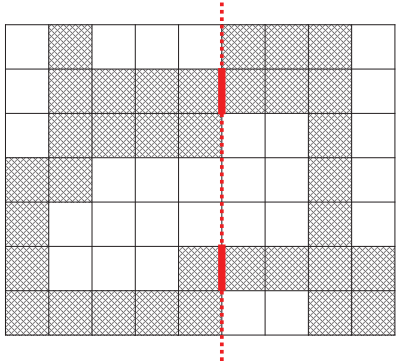}
  \caption{The cells of a domain $\Omega$ are diagonally hatched.
  The grid line $U'$ is denoted by a dotted red line.
  The domains $\Omega_1$ and $\Omega_2$ are to the left and right of  $U'$,
  respectively.
  The one-dimensional domain $U = \Omega_1 \cap \Omega_2$ is denoted by
  solid red line-segments.}
\label{figd2book2cut}
\end{figure}

\begin{lemma}
\label{d2stitching}
  Let a grid line $U'$ split a domain $\Omega$ into two domains:
  $\Omega =  \Omega_1 \cup \Omega_2$ and  $\Omega_1 \cap \Omega_2 = U$.
  Let $\mathcal{T}_1$, $\mathcal{T}_2$, and $\mathcal{T}$ be the T-meshes
  corresponding to $\Omega_1$, $\Omega_2$, and $\Omega$, respectively.
  For a given integers $m,n \geqslant 1$, suppose that the dimensions of
  the spaces $\mathcal{S}_{m,n} (\mathcal{T}_1)$ and $\mathcal{S}_{m,n} (\mathcal{T}_2)$
  can be obtained from \eqref{dimd2equ}.
  In addition, suppose that $U \in \mathcal{A}_{m-1} ^1$ (if $U$ is horizontal) and $U \in \mathcal{A}_{n-1} ^1$
  (if $U$ is vertical) with respect to the infinite one-dimensional grid $U'$.
  Then, the dimension of the spline space $ \mathcal{S}_{m,n} (\mathcal{T})$
  is given by \eqref{dimd2equ} as well.
\end{lemma}
\noindent {\it Proof.}
  Suppose that $U'$ is a vertical line $x = 0$.
  We can define the linear operator
  $$\mathcal{G}:
    \mathcal{S}_{m,n} (\mathcal{T}_1) \oplus \mathcal{S}_{m,n} (\mathcal{T}_2)
    \rightarrow \mathcal{S}_{n} (\mathcal{U})^{m} $$
  as follows: for given splines $\phi_1 \in \mathcal{S}_{m,n} (\mathcal{T}_1)$ and
  $\phi_2 \in \mathcal{S}_{m.n} (\mathcal{T}_2)$ the corresponding spline-vector
  $\mathcal{G}(\langle \phi_1, \phi_2 \rangle) \in  \mathcal{S}_{n} (\mathcal{U})^{m}$
  equals
  $$\big\langle (\phi_1 - \phi_2)|_{x=0},
    \frac{\partial (\phi_1 - \phi_2)}{\partial x}|_{x=0},\dots,
    \frac{\partial (\phi_1 - \phi_2)}{\partial x ^{m-1}}|_{x=0} \big\rangle, $$
  so $\mathrm{ker} \, \mathcal{G} = \mathcal{S}_{m,n} (\mathcal{T})$.
  Thus, we obtain
  \begin{equation}
  \begin{split}
    \mathrm{dim}\,\mathcal{S}_{m,n} (\mathcal{T}) =
    \mathrm{dim}\,\mathcal{S}_{m,n} (\mathcal{T}_1) +
    \mathrm{dim}\,\mathcal{S}_{m.n} (\mathcal{T}_2) -
    \mathrm{dim}\,\mathrm{im}\,\mathcal{G} = \\
    (m+1)(n+1) f_2 -  ( (m+1) n f_1 ^{h,0} +  (n+1) m (f_1 ^{v,0} - h_1 )) + \\
     m n (f_0 ^0 - h_0 ^0) -
     \mathrm{dim}\,\mathrm{im}\,\mathcal{G},
  \end{split}
  \end{equation}
  where $h_1$ and $h_0 ^0$ are the numbers of
  cells and inner vertices of the one-dimensional domain $U$.
  We remark that cells and inner vertices of $U$ are inner edges and
  inner vertices of $\Omega$, respectively, but are not inner edges and
  inner vertices of $\Omega_1$ and $\Omega_2$.
  Therefore, $f_1 ^{h,0}$, $f_1 ^{v,0} - h_1$ and $f_0 ^0 - h_0 ^0$ are the numbers of horizontal inner
  edges, vertical inner edges and inner vertices contained in either $\Omega_1$ or $\Omega_2$.

  In order to prove that $\mathcal{G}$ is an epimorphism, let us take
  an element of $\mathcal{S}_n (\mathcal{U})^{m}$: $\psi =
  \langle  \psi_1 (y), \dots, \psi_m (y)\rangle$.
  It follows from Corollary~\ref{extensiond1} that there exist
  splines $\widetilde{\psi}_1,\dots,\widetilde{\psi}_m$ defined globally
  over the infinite one-dimensional grid $U'$ such that
  $\widetilde\psi_i |_U = \psi_i , i =1 \dots m$.
  We define a bivariate spline $\phi (x,y)$ globally over $T'$ as follows:
  \begin{equation*}
   \phi (x,y) : = \sum_{i=1} ^{m} \widetilde{\psi}_i (y) \frac{x^{i-1}}{(i-1)!}  
  \end{equation*}
  Let $\phi_1 := \phi|_{\Omega_1}$ and $\phi_2 \equiv 0$ on $\Omega_2$.
  Then, $\mathcal{G} (\langle \phi_1, \phi_2 \rangle) =  \psi$.
  Thus, by virtue of Proposition~\ref{dimd1}, we obtain
  $\mathrm{dim} \, \mathrm{im} \, \mathcal{G} =
  m \, \mathrm{dim} \, \mathcal{S}_n (\mathcal{U}) =
  m ((n+1) h_1 -  n  h_0^0)$. For a horizontal grid line $U'$ the proof is analogous. Thus, the lemma is proved.  $\Box$
  \begin{corollary}
  \label{cordimd2}
  Let $\Omega \in \mathcal{A}_{m-1,n-1} ^2$ be a two-dimensional domain
  and $\mathcal{T}$ be the corresponding T-mesh.
  Then, the dimension of a space $\mathcal{S}_{m,n} (\mathcal{T})$ is
  \begin{equation}
  \label{findimformd2}
    \begin{split}
    \mathrm{dim}\, \mathcal{S}_{m,n} (\mathcal{T}) =
    (m+1) (n+1) f_2 -  ( (m+1) n  f_1 ^{h,0}  + \\
    (n+1) m f_1 ^{v,0} ) + m n f_0 ^0.
   \end{split}
  \end{equation}
\end{corollary}
\noindent {\it Proof.}
  Suppose that a domain $\Omega$ is split into two domains
  $\Omega_1$ and $\Omega_2$ by a vertical grid line $U'$ of $T'$
  (see Fig.~\ref{figd2book2cut}).
  From Proposition \ref{sectiondomd2} we obtain that $U \in \mathcal{A}_{n-1} ^1$.
  Using a sufficient number of vertical grid lines,
  a domain $\Omega$ can be split into pieces that are simply connected.
  By Lemma~\ref{d2stitching} and Proposition~\ref{cordimd2noholes},
  the corollary  is proved. $\Box$

\subsection{Basis of a spline space over a two-dimensional domain}
\label{subsecbasisd2}

  For a given integers $m,n \geqslant 1$, let $\mathcal{B}$ be the set of
  $(m+1) \times (n+1)$ rectangles formed by $(m+1)(n+1)$ cells of $T'$,
  so $\mathcal{B}$ is the set of all possible minimal supports for
  B-splines of bi-degree $m,n$ defined over $T'$ with knot multiplicities
  equal to $1$. We denote by $\widehat{\mathcal{B}}$ the collection of
  B-splines whose supports become the elements of $\mathcal{B}$.
  Let $\mathcal{N}$ be the number of elements of $\mathcal{B}$ that have
  at least one cell in common with a domain $\Omega$.
  \begin{proposition}
 \label{propnumbd2}
  Let $f_2$, $f_1 ^{h}$, $f_1 ^{h,0}$,$f_1 ^v$, $f_1 ^{v,0}$, $f_0$ and $f_0 ^0$ be the numbers of cells, horizontal edges,
  horizontal inner edges, vertical edges, vertical inner edges, vertices and inner vertices of a domain
  $\Omega \in \mathcal{A}_{0,0} ^2$.
  Then, the following identities hold:
  \begin{equation}
  \label{d2f1h}
    f_1 ^h = 2 f_2 -  f_1 ^{h,0} ,
  \end{equation}
  \begin{equation}
  \label{d2f1v}
    f_1 ^v = 2 f_2 -  f_1 ^{v,0} ,
  \end{equation}
  \begin{equation}
  \label{d2f0}
    f_0 = 4 f_2 - 2 (f_1 ^{h,0} +  f_1 ^{v,0} )+ f_0 ^0.
  \end{equation}
  \end{proposition}
  \noindent {\it Proof.}
  It is easy to see that the numbers of boundary horizontal and vertical edges are $2 f_2 - 2 f_1^{h,0}$
  and $2 f_2 - 2 f_1 ^{v,0}$, respectively.
  Thus, $2 f_2 - 2 f_1^{h,0} = f_1 ^h - f_1^{h,0}$ and $2 f_2 - 2 f_1^{v,0} = f_1 ^v - f_1^{v,0}$, which imply \eqref{d2f1h} and \eqref{d2f1v}.
  As long as $\Omega$ is a two-dimensional topological manifold with boundary,
  the boundary $\partial \Omega$ falls into piecewise linear curves that are
  connected, closed, and free of self-intersections.
  For each of these curves the number of edges is equal to
  the number of vertices.
  Thus, $(f_1 ^h + f_1 ^v) - (f_1^{h,0} + f_1 ^{v,0} )  = f_0 - f_0^0$, which implies \eqref{d2f0}. $\Box$
 \begin{lemma}
  For a given  couple of nonnegative integers
  $\overline{k}: = (k_1,k_2)$ let $\Omega \in \mathcal{A}_{k_1,k_2} ^2$.
  We denote by $f_{2,\overline{k}}$, $f_{1,\overline{k}} ^h$, $f_{1,\overline{k}} ^{h,0}$,  $f_{1,\overline{k}}
  ^v$, $f_{1,\overline{k}} ^{v,0}$,
  $f_{0,\overline{k}}$, $f_{0,\overline{k}} ^0$ the numbers of cells, horizontal edges, horizontal inner edges,
  vertical edges, vertical inner edges, vertices and inner vertices of the dilatation domain $\Omega_{k_1,k_2}^e$.
  Then, the following identities hold:
  \begin{equation}
  \label{d2f2k}
  \begin{split}
   f_{2,\overline{k}} = (k_1 +1) (k_2 +1) f_2 -  ( (k_1+1) k_2   f_1 ^{h,0}
   + \\    (k_2 + 1) k_1  f_1 ^{v,0} ) +     k_1  k_2 f_0 ^0,
  \end{split}
  \end{equation}
  \begin{equation}
  \label{d2f1kh}
   \begin{split}
   f_{1,\overline{k}} ^h =
   (k_1 + 1) (k_2 +2) f_2 -  ( (k_1+1) (k_2 + 1)  f_1 ^{h,0}
   + \\
   (k_2 + 2) k_1  f_1 ^{v,0} ) +    k_1  (k_2 + 1) f_0 ^0,
  \end{split}
  \end{equation}
  \begin{equation}
  \label{d2f1kv}
  \begin{split}
   f_{1,\overline{k}} ^v = (k_1 +2) (k_2 +1) f_2 -  ( (k_1+2) k_2   f_1 ^{h,0}
   + \\ (k_2 + 1) (k_1 +1) f_1 ^{v,0} ) +     (k_1 +1) k_2 f_0 ^0,
  \end{split}
  \end{equation}
  \begin{equation}
  \label{d2f0k}
  \begin{split}
   f_{0,\overline{k}}= (k_1 +2) (k_2 +2) f_2 -  ( (k_1+2) (k_2 + 1)  f_1 ^{h,0}
   + \\ (k_2 + 2) (k_1 +1) f_1 ^{v,0} ) +     (k_1 +1) (k_2 + 1) f_0 ^0.
  \end{split}
  \end{equation}
 \end{lemma}
 \noindent {\it Proof:}
  We will prove the theorem by induction on $k_1$ and $k_2$.
  For $k_1 = k_2 =0$, identity \eqref{d2f2k} is straightforward, and
  \eqref{d2f1kh},\eqref{d2f1kv} and \eqref{d2f0k} follow from
  \eqref{d2f1h},\eqref{d2f1v} and $\eqref{d2f0}$, respectively.
  Suppose that the theorem is proved for $\overline{k'} = (k_1-1, k_2)$.
  By Definitions~\ref{d2defdilatation} and~\ref{d2defoffset}, we have
  the following identities:
  \begin{equation}
  \label{d2identities1}
  f_{2,\overline{k}}  = f_{1,\overline{k'}} ^v \, , \, f_{1,\overline{k}} ^{h} = f_{0, \overline{k'}}  \, , \,  f_{1,\overline{k}} ^{v,0}  = f_{2,
  \overline{k'}}            \, , \, f_{0,\overline{k}} ^{0} = f_{1, \overline{k'}}
  ^{h,0} \, .
  \end{equation}
  Then, from \eqref{d2identities1} and Proposition \ref{propnumbd2} we finally
  have:
  \begin{equation}
  \label{d2identities2}
  f_{2,\overline{k}}  = f_{1,\overline{k'}} ^v \, , \, f_{1,\overline{k}} ^{h} = f_{0, \overline{k'}}  \, , \,
  f_{1,\overline{k}} ^{v}  =  2 f_{1,\overline{k'}} ^v - f_{2, \overline{k'}}   \, , \, f_{0,\overline{k}}  =
  2 f_{0, \overline{k'}} - f_{1, \overline{k'}} ^h \,.
  \end{equation}
  From the supposition that \eqref{d2f2k}--\eqref{d2f0k} hold for $\overline{k'}= (k_1 -1,
  k_2)$ and the identities \eqref{d2identities2} we obtain that
  \eqref{d2f2k}--\eqref{d2f0k} hold for $\overline{k}= (k_1, k_2)$.
  Similarly, suppose that the theorem is proved for $\overline{k''} =
  (k_1, k_2 -1)$.
  By Definitions~\ref{d2defdilatation} and~\ref{d2defoffset}, we have
  the following identities:
  \begin{equation}
  \label{d2identities3}
  f_{2,\overline{k}}  = f_{1,\overline{k''}} ^h \, , \, f_{1,\overline{k}} ^{v} = f_{0, \overline{k''}}  \, , \,  f_{1,\overline{k}} ^{h,0}  = f_{2,
  \overline{k''}} \, , \, f_{0,\overline{k}} ^{0} = f_{1, \overline{k''}} ^{v,0} \, .
  \end{equation}
  Then, from \eqref{d2identities3} and Proposition \ref{propnumbd2} we finally
  have:
  \begin{equation}
  \label{d2identities4}
  f_{2,\overline{k}}  = f_{1,\overline{k''}} ^h \, , \, f_{1,\overline{k}} ^{v} = f_{0, \overline{k''}}  \, , \,
  f_{1,\overline{k}} ^h  = 2 f_{1,\overline{k''}} ^h - f_{2, \overline{k''}} \, , \, f_{0,\overline{k}}  =
  2 f_{0, \overline{k''}} - f_{1, \overline{k''}} ^v \, .
  \end{equation}
  From the supposition that \eqref{d2f2k}--\eqref{d2f0k} hold for
  $\overline{k''}= (k_1 , k_2 - 1)$ and the identities \eqref{d2identities4} we obtain that
  \eqref{d2f2k}--\eqref{d2f0k} hold for $\overline{k}= (k_1, k_2)$.$\Box$

  \begin{corollary}
  \label{cord2}
  Suppose that $\Omega \in \mathcal{A}_{m-1,n-1} ^2$.
  Then, the following identity holds:
  \begin{equation}
  \label{d2N}
  \begin{split}
    \mathcal{N} = (m+1) (n+1) f_2 -  ( (m+1) n  f_1 ^{h,0}  + \\
    (n+1) m f_1 ^{v,0} ) + m n f_0 ^0 .
  \end{split}
  \end{equation}
  \end{corollary}
  \noindent {\it Proof.}
  Each $(m+1) \times (n+1)$ square from $\mathcal{B}$ is associated with
  its centroid.
  If $m$ and $n$ are odd, then this centroid is a grid node of $T'$; if $m$ is even and $n$ is odd,
  then this centroid is a grid node of $T'_{1,0}$; if $m$
  is odd and $n$ is even, then this centroid is a grid node of
  $T'_{0,1}$; if $m$ and $n$ are even, then this centroid
  is a grid node of $T'_{1,1}$.

  It can be seen that an element of $\mathcal{B}$ has at least one cell in common
  with $\Omega$ iff its centroid is a vertex of
  the dilatation domain $\Omega_{m-1,n-1} ^e$.
  Thus, $\mathcal{N}=f_{0,\overline{k}}$ for $\overline{k} = (m-1,n-1)$  and from \eqref{d2f0k} we obtain \eqref{d2N}. $\Box$

 \begin{corollary}
 \label{maintheoremd2}
  For a given couple of integers $m,n \geqslant 1$, suppose that $\Omega \in \mathcal{A}_{m-1,n-1}^2$.
  Then, the basis of a space $\mathcal{S}_{m,n} (\mathcal{T})$ can be obtained
  as follows:
  \begin{equation*}
   \{ b|_{\Omega} : b (x,y) \in \widehat{\mathcal{B}} \,\, \wedge \,\,
   \mathrm{supp} \, b (x,y) \cap \mathrm{int} \, \Omega\neq\varnothing \}.
  \end{equation*}
\end{corollary}
\noindent{\it Proof.}  Corollary \ref{maintheoremd2} is a direct
consequence  of Corollaries~\ref{cordimd2} and~\ref{cord2}, and the
fact that tensor--product B-splines $\widehat{\mathcal{B}}$ are
linear independent. $\Box$

\section{Hierarchical splines}
\label{hiersection}
  
 In this section we mostly follow the notations introduced by Giannelli and J\"{u}ttler in \cite{Carlotta11}.

 For a given integers $m,n \geqslant 1$, let $ V^0 \subset V^1 \subset \dots \subset V^{N-1} $ be a nested sequence of
 $N$ spaces of bivariate splines of bi-degree $(m,n)$, with the knot multiplicities equal to
 $1$, associated with a sequence of infinite two-dimensional grids $G^0 \subset G^1 \subset \dots \subset G^{N-1}$.
 We denote by $T^{\ell}$ a tensor-product B-spline basis that spans the spline space $V^{\ell},  \ell=0,\dots,N-1$.
 Let us consider a nested sequence of domains $\Omega^0 \supset
 \Omega^1 \supset \dots \supset \Omega^{N-1} \supset \Omega^N =
 \varnothing$ such that each domain $\Omega^\ell$ is formed by a finite
 number of cells of $G^{\ell}, \ell=0,\dots,N-1$. We require that for each
 $\ell=1,\dots,N-1$ the boundary $\partial \Omega^\ell$ is aligned with
 grid lines of $G^{\ell-1}$.
 Let $\mathcal{H}$ be the T-mesh determined by a nested
 sequence of domains $\Omega^0 \supset \Omega^1 \supset \dots \supset \Omega^{N-1} \supset \Omega^N = \varnothing$ associated with a
 nested sequence of grids $G^0 \subset G^1 \subset \dots \subset G^{N-1}$. Let $\mathcal{S}_{m,n} (\mathcal{H})$ be the space of splines of bi-degree
 $(m,n)$, with maximal order of smoothness, defined over the  hierarchical T-mesh $\mathcal{H}$.
 Fig.~\ref{hiermeshfig} (left) shows a simple example of a hierarchical T-mesh $\mathcal{H}$ determined by the nested
  sequence of two-dimensional domains $\Omega^0 \supset \Omega^1 \supset \Omega^2 \supset \Omega^3 =
  \varnothing$.
 \begin{figure}[htp] \centering
  \includegraphics[width=1 \textwidth]{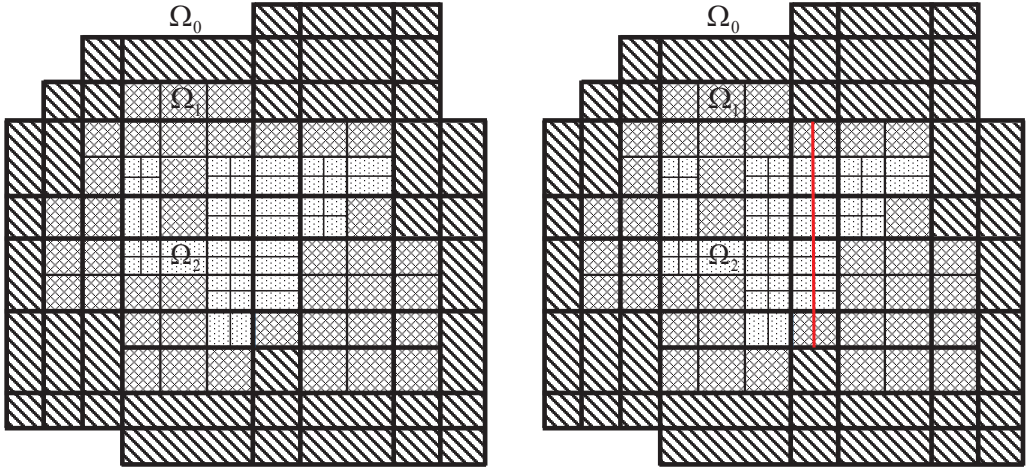}
  \caption
  {
    A T-mesh $\mathcal{H}$ (left) is determined by domains $\Omega^0 \supset \Omega^1 \supset \Omega^2
    $: $\Omega^0$ is all the hatched area, $\Omega^1$ is an union of cross-diagonally shaded area and dotted area,
    $\Omega^2$ is a dotted area.
    In order to obtain a T-mesh $\mathcal{H}_{l_1}^1$ (right), the T-mesh
    $\mathcal{H}$ is refined by the line-segment $l_1 \cap \Omega^1$ (the red line-segment).
  }
\label{hiermeshfig}
\end{figure}

The hierarchical B-splines are defined through the following
selection mechanism:
 \begin{definition}[\cite{Carlotta11}, Definition~1]
 \label{kraftbasisdef}
  The set of hierarchical B-splines  $\mathcal{K}$ is defined as
  \begin{equation*}
    \mathcal{K} = \bigcup\limits_{l=0}^{N-1} \mathcal{K}^\ell,
  \end{equation*}
  with $\mathcal{K}^{\ell} = \{\tau \in T^\ell : \mathrm{supp} \,\tau\bigcap \, \mathrm{int} \, R^{\ell-1}
  = \varnothing \,\, \wedge \,\, \mathrm{supp} \,\tau \bigcap \, \mathrm{int} \, R^\ell
  \neq \varnothing \}$, where $R^\ell = \Omega^0 \setminus \Omega^{\ell+1}$ for $\ell= 0 \dots N-1$.
  We define $R^{-1} = \varnothing$ to include the case $\ell=0$.
\end{definition}
  As a consequence of Corollary \ref{maintheoremd2} we obtain Theorem \ref{finalcarlotta}:
  \begin{theorem} 
  \label{finalcarlotta}
  For a given nested sequence of domains $\Omega^0 \supset \Omega^1 \supset \dots \supset \Omega^{N-1} \supset \Omega^N = \varnothing$
  associated with the nested sequence of grids $G^0 \subset G^1 \subset \dots \subset G^{N-1}$,
  suppose that the domain $R^\ell = \Omega^0 \setminus \Omega^{\ell+1} \in \mathcal{A}_{m-1,n-1}^2 $
  with respect to the grid  $G^\ell$  for each $\ell= 0,\dots,N-1$.
  Then, the set of B-splines from $\mathcal{K}$ restricted on $\Omega^0$ forms a basis of  the spline space $\mathcal{S}_{m,n}(\mathcal{H})$.
 \end{theorem}
 \noindent {\it Proof:}   
  The theorem follows directly from
  Corollary \ref{maintheoremd2}.
   Indeed, the linear independence of B-splines from $\mathcal{K}$ is
  a trivial observation due to the linear independence of
  tensor-product B-splines.

  Let us prove that B-splines from $\mathcal{K}$ span $\mathcal{S}_{m,n}(\mathcal{H})$.
  Let $f \in \mathcal{S}_{m,n}(\mathcal{H})$. Since $R^0 \in \mathcal{A}_{m-1,n-1} ^2$ with respect to the grid  $G^0$, there exists
  $f^0 = \sum\limits_{b \in \mathcal{K}^0} c_b ^0 b \in
  V^0$, for proper real numbers $c_b^0, b \in \mathcal{K}^0$, such that $f|_{R^0} = f^0 |_{R^0}$. 
  Since $R^1 \in \mathcal{A}_{m-1,n-1} ^2$ with respect to the grid $G^1$, there exists $f^1 = \sum\limits_{b \in T^1} c_b  ^1 b \in V^1$
  such that $f^1 |_{R^1} = \left( f - f^0 \right)|_{R^1}$. Since $f^1|_{R^0}=0$, we obtain $c_b ^1=0$
  for any $b \notin \mathcal{K}^1$, and thus $f^1 = \sum\limits_{b \in \mathcal{K}^1} c_b ^1 b$.
 Repeating this procedure, we obtain $f = \sum\limits_{\ell=0}^{N-1} f^{\ell}|_{\Omega_0}$
 such that $f^{\ell} |_{R^\ell} = f|_{R^\ell} - \sum\limits_{i=0}^{\ell-1} f^i |_{R^\ell} $
 and $f^\ell = \sum\limits_{b \in \mathcal{K}^\ell} c_b ^\ell b$
  for $\ell = 0, \dots , N-1$. For more detailed proof we refer the reader to \cite{Carlotta11}, Theorem 20.$\Box$

  Theorem \ref{finalcarlotta} shows that if a T-mesh $\mathcal{H}$
  is in a suitable class, then hierarchical B-splines from
  $\mathcal{K}$ generates all basis functions of the spline space $\mathcal{S}_{m,n}(\mathcal{H})$.

   In the definition below we will introduce the basic iteration for a refinement of a hierarchical T--mesh. In particular, 
   we will need this definition to prove Corollary \ref{sumtoone} where the sufficient condition for 
   a collection of hierarchical B-splines to form a partition of unity is given. 
   \begin{definition}
  \label{defrefofhier}
  Let $\mathcal{H}$ be a hierarchical T-mesh determined by a nested sequence of
  domains $\Omega^0 \supset \Omega^1 \supset \dots \supset \Omega^{N-1} \supset \Omega^N =
  \varnothing$ associated with a nested sequence of grids
  $G^0 \subset G^1 \subset \dots \subset G^{N-1}$.
  For a given line $l_1 \not\subset G^{N-1}$ and $0 \leqslant j_1 \leqslant N-1$, we denote by $\mathcal{H}_{l_1} ^{j_1}$ the T-mesh determined by the
  sequence of domains $\Omega^0 \supset \dots  \supset \Omega^{j_1-1} \supset \Omega^{j_1} \supset  \dots \supset \Omega^{N-1} \supset \Omega^N =
  \varnothing$ associated the nested sequence of grids $G^0 \subset \dots \subset G^{j_1-1} \subset G^{j_1} \cup \{l_1 \}
  \subset \dots \subset G^{N-1} \cup \{l_1\}$.
  Fig.~\ref{hiermeshfig} (right) shows an example of the refined T-mesh $\mathcal{H}_{l_1} ^1$.
  We note that if $l_1 \cap \mathrm{int}\, \Omega_{j_1} = \varnothing$,
  then $\mathcal{H}_{l_1} ^{j_1} = \mathcal{H}$.
  We denote by
  $\mathcal{H}_{l_1,\dots, l_k} ^{j_1, \dots ,j_k}$ the T-mesh that
  is obtained from $\mathcal{H}$ by consecutive refinements by the
  lines $l_1, \dots , l_k$ at levels $j_1,\dots, j_k$, respectively.
  \end{definition}

  In Proposition \ref{refhierprop1} below we will show that the condition of Theorem \ref{finalcarlotta}
  holds true if one continues
   to refine $\mathcal{H}$ in a way given in Definition \ref{defrefofhier}.
   \begin{proposition}
  \label{refhierprop1}
  Let $\mathcal{H}$ be a T-mesh determined by a nested sequence of
  domains $\Omega^0 \supset \Omega^1 \supset \dots \supset \Omega^{N-1} \supset \Omega^N =
  \varnothing$ associated with a nested sequence of grids
  $G^0 \subset G^1 \subset \dots \subset G^{N-1}$.
  Suppose that the conditions
  of Theorem \ref{finalcarlotta} are fulfilled, i.e. each domain $R ^\ell = \Omega^0 \setminus
  \Omega^{\ell+1}$, considered with respect to the grid $G^\ell$, belongs to the class
  $\mathcal{A}_{m-1,n-1}^2$ for any $\ell =0 \dots N-1$. Then, for  the
  refined T-mesh $\mathcal{H}_{l_1, \dots , l_k} ^{j_1, \dots,
  j_k}$ the conditions of Theorem \ref{finalcarlotta} are fulfilled as well.
  \end{proposition}
  \noindent {\it Proof.}
   Let us prove Proposition \ref{refhierprop1} for $k=1$.
   For the refined T-mesh $\mathcal{H}_{l_1} ^{j_1}$, the domains $\Omega^0 \supset \dots  \supset
   \Omega^{N-1}$ and the grids $G^0 \subset \dots \subset G^{j_1-1}$ remain untouched, but the the grids
   $G^{j_1} \subset \dots \subset G^{N-1}$ change to $G^{j_1} \cup \{l_1\} \subset \dots \subset G^{N-1} \cup
   \{l_1\}$, respectively.
   Thus, we need to prove that the domains $ R^0 \subset \dots \subset R^{j_1 -1} \subset R^j \subset \dots \subset R^{N-1} = \Omega^0$,
   considered with respect to the grids $G^0 \subset \dots \subset G^{j_1 -1} \subset G^{j_1} \cup \{ l_1 \} \subset \dots \subset
   G^{N-1} \cup \{l_1 \}$ respectively, belong to the class $\mathcal{A}_{m-1,n-1}
   ^{2}$ as well.
   For $R^\ell, \ell < j_1$ there is nothing to prove.
   For $R^\ell,\ell \geqslant j_1$ the proof follows from Proposition
   \ref{newlined2prop}. The proof of Proposition \ref{refhierprop1} for $k>1$ follows by induction on $k$.$\Box$

 Proposition \ref{finalcarlottaprop} below provides us a sufficient condition to apply
 Theorem \ref{finalcarlotta}; the condition is expressed in terms of domains $\Omega^0, \dots ,
 \Omega^{N-1}$ themselves rather than difference sets $\Omega^0 \setminus \Omega^{\ell+1}, \ell= 0 \dots
 N-1$.
 \begin{proposition}
 \label{finalcarlottaprop}
    Let $\mathcal{H}$ be a T-mesh determined by  $\Omega^0 \supset \Omega^1 \supset \dots \supset \Omega^{N-1} \supset \Omega^N =
    \varnothing$  associated with the nested sequence of grids $G^0 \subset G^1 \subset \dots \subset G^{N-1}$.
    Suppose that $\Omega^0 \in \mathcal{A}_{m-1,n-1}^2$ with
    respect to $G^0$ and $\Omega^\ell \in    \widetilde{\mathcal{A}}_{m-1,n-1}^2$ with respect to the grid $G^{\ell-1}$ for $ \ell= 1 \dots N-1$. In
    addition, suppose that $\partial \Omega^0  \cap \partial \Omega^1 =\varnothing$.
    Then, the conditions of Theorem \ref{finalcarlotta} are fulfilled.
 \end{proposition}
 \noindent {\it Proof.} An assumption $\partial \Omega^0  \cap \partial \Omega^1
 =\varnothing$ implies that $\Omega^\ell \subset  \mathrm{int} \, \Omega^0$ for $\ell =1 \dots
 N-1$. Since $\Omega^0 \in \mathcal{A}_{m-1,n-1}^2$ with respect $G^0$, it follows from Corollary \ref{newlined2} that
 $\Omega^0 \in \mathcal{A}_{m-1,n-1} ^2$ with respect to $G^\ell$ for $\ell=0 \dots N-1$.
 Since $\Omega^{\ell+1} \in \widetilde{\mathcal{A}}_{m-1,n-1} ^2$ with respect to $G^\ell$ and $\Omega^{\ell+1} \subset
 \mathrm{int}\,\Omega^0$, then
 $R^\ell = \Omega^0 \setminus \Omega^{\ell+1} \in \mathcal{A}_{m-1,n-1} ^2$ with respect to $G^\ell$ for $\ell =0 \dots
 N-1$. Thus, the conditions of Theorem \ref{finalcarlotta} are fulfilled.$\Box$

  Proposition \ref{finalteor2} below will be used for the proof of
  Corollary \ref{sumtoone} where we show that under the certain condition on a hierarchical
  T-mesh hierarchical B-splines provide a weighted partition of unity
  for some positive weights.

 \begin{proposition}
 \label{finalteor2}

    Let $\mathcal{H}$ be a T-mesh determined by  $\Omega^0 \supset \Omega^1 \supset \dots \supset \Omega^{N-1} \supset \Omega^N =
    \varnothing$  associated with the nested sequence of grids $G^0 \subset G^1 \subset \dots \subset G^{N-1}$.
    Let 
    $\Omega^0 \in \mathcal{A}_{m-1,n-1}^2$ with
    respect to $G^0$ and $\Omega^\ell \in    \widetilde{\mathcal{A}}_{m-1,n-1}^2$ with respect to the grid $G^{\ell-1}$ for $ \ell= 1 \dots
    N-1$, and  $\partial \Omega^0  \cap \partial \Omega^1 =\varnothing$.
    In addition, we suppose that $\Omega^\ell \in  \widetilde{\mathcal{A}}_{m,n}^2$ with respect to the grid $G^\ell$ for $\ell =1 \dots N-1$.
    We denote by  $\mathcal{K} (\mathcal{H}) = \bigcup\limits_{\ell=0} ^{N-1} \mathcal{K}^\ell (\mathcal{H})$  and
    $\mathcal{K} (\mathcal{H}_{l_1,\dots, l_k} ^{j_1,\dots,j_k}) = \bigcup\limits_{\ell=0}^{N-1} \mathcal{K}^\ell (\mathcal{H}_{l_1,\dots,l_k}
    ^{j_1,\dots,j_k})$  the sets of hierarchical B-splines
    given in Definition \ref{kraftbasisdef} for  T-meshes $\mathcal{H}$
    and $\mathcal{H}_{l_1,\dots,l_k} ^{j_1,\dots,j_k}$, respectively.
    If the collection of hierarchical B-splines $\mathcal{K}(\mathcal{H})$
    provides a weighted partition of unity  $\sum_{\tau \in \mathcal{K} (\mathcal{H})} w_{\tau} ^0   \tau |_{\Omega^0} =
    1$ for some positive weights $w_{\tau}^0, \tau \in \mathcal{K}
    (\mathcal{H})$, then the collection of hierarchical B-splines $\mathcal{K}(\mathcal{H}_{l_1,\dots,l_k}
    ^{j_1,\dots,j_k})$ provides a weighted partition of unity
    $\sum_{\tau \in \mathcal{K} (\mathcal{H}_{l_1,\dots,l_k} ^{j_1,\dots,j_k})} w_{\tau} ^1  \tau |_{\Omega^0} =  1$ for some positive
    weights $w_{\tau} ^1 , \tau \in \mathcal{K} (\mathcal{H}_{l_1,\dots,l_k} ^{j_1,\dots,j_k})$ as well.

 \end{proposition}
 \noindent {\it Proof.}
   Before proving Proposition \ref{finalteor2} let us note that from  Propositions \ref{refhierprop1},\ref{finalcarlottaprop}
   and Theorem \ref{finalcarlotta} we have that the hierarchical B-splines from
   $\mathcal{K}(\mathcal{H})$ and $\mathcal{K} (\mathcal{H}_{l_1,\dots,l_k} ^{j_1,\dots,j_k})$, restricted on the domain $\Omega^0$, form bases of the
   spaces $\mathcal{S}_{m,n}(\mathcal{H})$ and $\mathcal{S}_{m,n} (\mathcal{H}_{l_1,\dots,l_k} ^{j_1,\dots,j_k})$, respectively.
   We will prove Proposition \ref{finalteor2} for the basic case
   $k=1$; for $k>1$ it follows by induction on $k$.

  In order to prove Proposition \ref{finalteor2}, we need to show that for any $\tau \in \mathcal{K}^\ell (\mathcal{H}_{l_1}
  ^{j_1}), \ell \geqslant j_1 $ there exits
  $\tau' \in \mathcal{K}^\ell (\mathcal{H})$ such that $\mathrm{supp}\, \tau \subset  \mathrm{supp} \
  \tau'$.\footnote{This condition is also mentioned  by Voung et al.\cite{Vuong11}, see the subsection 2.5; 
  in that subsection the partition of unity property for a weighted hierarchical B--splines is discussed under more general settings.}       We suppose that $\mathrm{supp} \, \tau  \cap (\Omega^{\ell} \cap l_1) \neq
  \varnothing$; otherwise, $\tau '$ simply equals to $\tau$.
  If $\tau \in \mathcal{K} ^0 (\mathcal{H}_{l_1} ^{j_1})$, then, by Definition
  \ref{kraftbasisdef},
  $\mathrm{supp}\, \tau \cap \mathrm{int} \, R^0 \neq \varnothing $.
  Obviously, there exists $\tau' \in T^0$
  such that $\mathrm{supp} \, \tau \subset \mathrm{supp} \, \tau' $,
  therefore $\mathrm{supp}\, \tau' \cap \mathrm{int} \, R^0 \neq
  \varnothing$ which implies that $\tau' \in \mathcal{K} ^0
  (\mathcal{H})$. Hereinafter we will assume that $\tau \in \mathcal{K}^\ell (\mathcal{H}_{l_1}
  ^{j_1})$ for $\ell \geqslant 1$.
  The support $B:= \mathrm{supp} \, \tau$ is a rectangle formed by $(m+1) \times (n+1) $ cells of the grid $G^\ell \cup {l_1}$,
  where  $B \subset \Omega^\ell$. Thus, we need to show that there exists a rectangle $B' \subset \Omega^\ell$ formed by $(m+1) \times (n+1)$
  cells of the grid $G^\ell$ such that $B \subset B'$.

  \begin{figure}[htp] \centering
  \includegraphics[width=0.5 \textwidth]{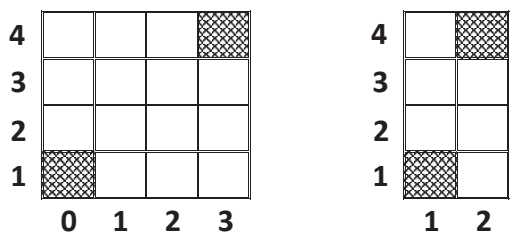}
  \caption
  {A rectangle $B''$ (left) is shown for the case $m=2$ and $n=3$. The diagonally hatched cells (corresponding to the indices $i_1 = 1$
  and $i_2 = 4$) belong to a domain $\widetilde{\Omega}^\ell$. The diagonally hatched cells in a rectangle $C$ (right)
  belong to a domain $(\widetilde{\Omega}^\ell)_{2,0} ^e$.}
  \label{book1d}
  \end{figure}

  Let us assume that $l_1$ is a vertical line.
  If $B \cap l_1 = \partial B \cap l_1$, i.e. one side of the rectangular
  boundary of $B$ is aligned with $l_1$, then there is nothing to prove since
  the desired rectangle $B'$ always exists due to the condition that $\partial \Omega^\ell
  $ is aligned with the lines of the grid $G^{\ell-1} \subset G^{\ell}$.
  If $l_1$ splits $B$, then $B$ is a rectangle formed by $m \times (n+1)$ cells of the grid $G^\ell$.
  We denote by $B''$ a rectangle formed by $(m+2) \times (n+1)$ cells of the grid $G^\ell$ that contains $B$ and two columns of cells
  from the left and right side of $B$.
  Let us index the columns of $B''$ by the numbers $0, \dots, m+1 $
  from the left to the right and rows by the numbers $1, \dots, n+1$ from the bottom to
  the top. Then, each cell from $B$ is indexed by a pair $(i,j)$,
  where $1 \leqslant i \leqslant n+1$ and $1 \leqslant j \leqslant m$ are the indices of row
  and column respectively that contain this cell. The cells of the column from the
  left and from the right side of $B$ are indexed by pairs $(i,0)$ and $(i,m+1)$ respectively, where $1 \leqslant i \leqslant
  n+1$. We denote the left and right column by $C_1$  and $C_2$
  respectively, i.e. $B'' =  B \cup C_1 \cup C_2$.
  By Definition \ref{d2inneroffset} we have that $\widetilde{\Omega} ^\ell  := \widehat{\Omega} \setminus
  \Omega^\ell \in \mathcal{A}_{m,n} ^2$ with respect to the grid $G^\ell$,
  where $\widehat{\Omega}$ is a sufficiently large rectangle
  that contains $\Omega^\ell$ in its interior $\mathrm{int}\,
  \widehat{\Omega}$.

  Let us prove by contradiction that either $C_1 \subset \Omega^\ell$ or $C_2 \subset \Omega^\ell$. Assume that there
  exist at least one cell in $C_1$ and one cell in $C_2$ that
  does not belong to $\Omega^\ell$. From this assumption we have that
  $\widetilde{\Omega} ^\ell$ has at least one common cell with each of the
  columns $C_1$ and $C_2$. Since $\widetilde{\Omega} ^\ell \in
  \mathcal{A}_{m,n}^2$ with respect to the grid $G^\ell$, then $\widetilde{\Omega}  ^\ell \in
  \mathcal{A}_{m,0}^2$ with respect to the same grid. From Definition \ref{d2defoffset} 
   one may conclude that there exist $i_1, i_2$,
  where $1 \leqslant i_1, i_2 \leqslant n+1$ and $|i_1 - i_2| > 1$,
  such that the cells with the indices $(i_1,0)$ and $(i_2,m+1)$ belongs
  to $\widetilde{\Omega}^\ell$, but the cell with the indices $(i,0)$, for $i \in ( i_1, i_2]$, and
  $(i,m+1)$, for $i \in [ i_1, i_2 )$, do not belong to $\widetilde{\Omega}^\ell$.

  Let us take the dilatation domain $(\widetilde{\Omega}^\ell)_{m,0}^e$. This domain is formed by the cells of the
  grid $G_{m\,\mathrm{mod}\,2\,,\, 0}^\ell$.
  Also, it follows from Definition \ref{d2defdilatation} that there exists a rectangle $C$
  formed by $2 \times (n+1)$ cells of the $G_{m\,\mathrm{mod}\,2\,,\, 0}^\ell$
  (with the columns and rows indexed by the numbers $1,2$ and
  $1,\dots,n+1$, respectively) with following property: a cell of $C_1$ with the index $(i,0)$ belongs to $\widetilde{\Omega}^\ell$
  iff the cell of $C$ of the index $(i,1)$ belongs  $(\widetilde{\Omega}^\ell)_{m,0}^e$, and
  a cell of $C_2$ with the index $(i,m+1)$ belongs to $\widetilde{\Omega}^\ell$ iff the cell of $C$ of the index $(i,2)$
  belongs  $(\widetilde{\Omega}^\ell)_{m,0}^e$. From Definition \ref{d2defoffset} we know that $(\widetilde{\Omega}^\ell)_{m,0}^e$, considered with respect to
  $G_{m\,\mathrm{mod}\,2\,,\, 0}^\ell$, belongs to the class $\mathcal{A}_{0,n}^2$. But, since $|i_1 - i_2| \leqslant n$, we
  get that $ (\widetilde{\Omega}^\ell)_{m,0}^e \notin \mathcal{A}_{0,n} ^2$, which implies that  $\widetilde{\Omega}^\ell \notin \mathcal{A}_{m,n} ^2$.
  Thus, we have a contradiction. Therefore, either $C_1 \subset \Omega^\ell$ or $C_2 \subset \Omega^\ell$.
  So $B'$  might be chosen either as  $B \cup C_1$ or $B \cup C_2$.
  Fig.~\ref{book1d} shows the example for the case $m=2$ and $n=3$, and the indices $i_1 = 1$ and $i_2 = 4$.
  For a horizontal line $l_1$ the proof could be given in an analogous way.$\Box$

 \begin{corollary}
 \label{sumtoone}
    Let $\mathcal{H}$ be a T-mesh determined by  $\Omega^0 \supset \Omega^1 \supset \dots \supset \Omega^{N-1} \supset \Omega^N =
    \varnothing$  associated with the nested sequence of grids $G^0 \subset G^1 \subset \dots \subset G^{N-1}$.
    Suppose that $\Omega^0 \in \mathcal{A}_{m-1,n-1}^2$ with respect to
    $G^0$, and $\Omega^\ell \in    \widetilde{\mathcal{A}}_{m,n}^2$ with respect to the grid $G^{\ell-1}$ for $ \ell= 1 \dots
    N-1$. In addition, we suppose that $\partial \Omega^0 \cap \partial \Omega^1 =
    \varnothing$. Then, a collection of hierarchical B-splines $\mathcal{K}(\mathcal{H})$
    provides a weighted partition of unity  $\sum_{\tau \in \mathcal{K} (\mathcal{H})} w_{\tau} \tau |_{\Omega^0} =  1$ for some positive
    weights $w_{\tau} , \tau \in \mathcal{K} (\mathcal{H})$.
 \end{corollary}
 \noindent {\it Proof.}
  Let $\mathcal{H}^0$ be a tensor-product mesh determined by $\Omega^0 \supset
  \varnothing$ associated with the grid $G^0$. Then, $\sum_{\tau \in \mathcal{K} (\mathcal{H}^0)} \tau|_{\Omega^0}  =1$.
  It can be seen that $\mathcal{H}$ may be constructed from $\mathcal{H}^0$ by successive
  refinements in a way given in Definition
  \ref{defrefofhier}. Thus, Corollary \ref{sumtoone} is a direct
  consequence of Proposition \ref{finalteor2}.$\Box$

\section{Remarks}
\label{remarksection}

    \begin{remark}
    It can be seen that for $k_1=k_2 =k$ the class $\mathcal{A}_{k,k} ^2$ given in
    Definition \ref{d2defoffset} coincides with the class $\mathcal{A}_k ^2$ given in \cite{Berd13} (see Definition 3).
    The description of the basic class $\mathcal{A}_1 ^2$ \cite{Berd13} (see Definition 4)
    coincides with the one given in the original paper \cite{Carlotta11}, where the
    admissible types of intersections between a domain and the cells from
    the offset region are shown.
   \end{remark}

   \begin{remark}
     In Corollary \ref{sumtoone} the condition that $\Omega^\ell \in \widetilde{\mathcal{A}}_{m,n}^2$
     with respect to the grid $G^{\ell-1}$ for  $\ell=1,\dots,N-1$ cannot be simply weakened.
     Under the more general condition on a hierarchical T-mesh, given
     in Theorem \ref{finalcarlotta}, a weighted partition of unity
     cannot be always achieved for only positive weights.
     However, partition of unity property always holds for the modified hierarchical B-splines, namely THB-splines
    introduced in \cite{Carlotta12}.
\end{remark}

   \begin{remark}
   \label{essentialremark2}
    
     We note that Definition \ref{d2defoffset}, 
     Corollaries \ref{cordimd2}, \ref{cord2}, \ref{maintheoremd2}, \ref{sumtoone} and Theorem \ref{finalcarlotta}
     can be straightforwardly extended for the $d$--variate case $(m_1,\dots,m_d)$ for an arbitrary $d > 2$.
   
 \end{remark}

\section{Addendum} 
 The preprint version of this paper appeared in May 2013.  
 In October 2013, Mokri\v{s}, J\"{u}ttler and Giannelli published the preprint 
 \cite{Mokris13} where they extended the results of \cite{Carlotta11} and, in addition, provided new insight for some of the results from \cite{Berd13}.  
  We note that Theorem \ref{finalcarlotta}  can be alternatively obtained as a consequence of Theorem 3.5~\cite{Mokris13}; 
  it follows from the observation that the restriction on the configuration of domains given in Theorem 3.5~\cite{Mokris13} 
  is weaker than that of in Theorem \ref{finalcarlotta}.
 
\bibliographystyle{elsarticle-num}

\bibliography{refhier}

\end{document}